\documentclass[aps,prl,preprint,superscriptaddress]{revtex4}
\usepackage{graphics}

\usepackage{dcolumn}
\usepackage{bm}

\def\be{\begin{equation}}
\def\ee{\end{equation}}
\def\bea{\begin{eqnarray}}
\def\eea{\end{eqnarray}}

\begin{document}
\title{Analysis of the paraconductivity in underdoped $La_{2-x}Sr_xCuO_4$ thin films using high magnetic fields}
\author{B. Leridon}
\affiliation{CNRS/UPR5, ESPCI , 10  rue Vauquelin, 75231 Paris cedex 05, France}
\author{ J. Vanacken}, 
\affiliation{INPAC, Institute for Nanoscale Physics and Chemistry, Katholieke Universiteit Leuven, Celestijnenlaan 200 D, B-3001 Heverlee, Belgium}
\affiliation{ LNCMP, 143 avenue de Rangueil, 31400 Toulouse, France}

\author{ T. Wambecq}
\affiliation{INPAC, Institute for Nanoscale Physics and Chemistry, Katholieke Universiteit Leuven, Celestijnenlaan 200 D, B-3001 Heverlee, Belgium}

\author{V.V. Moshchalkov}
\affiliation{INPAC, Institute for Nanoscale Physics and Chemistry, Katholieke Universiteit Leuven, Celestijnenlaan 200 D, B-3001 Heverlee, Belgium}

\begin{abstract}
The contribution of superconducting fluctuations to the conductivity, or paraconductivity is studied in the underdoped regime of $La_{2-x}Sr_xCuO_4$ cuprates.  A perpendicular magnetic field up to 50 T is applied to suppress the superconductivity and obtain the normal state resistivity which is then used to calculate the paraconductivity. Surprisingly enough, it is consistent with a two-dimensional Aslamazov-Larkin (AL) regime of Gaussian fluctuations close to the critical temperature.  At higher temperature, the paraconductivity shows a power-law decrease in temperature (as $T^{-\alpha}$) as was previously shown for underdoped $YBa_2Cu_3O_{7-\delta}$ and $Bi_2Sr_2CaCu_2O_{8+\delta}$  samples. Our observations are not consistent with the existence of Kosterlitz-Thouless fluctuations. This tends to indicate that the superconducting pair amplitude is not already defined above $T_C$  in the pseudogap state.

\end{abstract}

\maketitle

\section{introduction}

High-Tc cuprate superconductors are known to exhibit a depression in the density of states, often referred to as the "pseudogap".  This feature, firstly discovered by NMR \cite{Alloul:1989} and also observed in the specific heat \cite{Liang:1996} is seen to occur below a given temperature $T*$ in the underdoped part of the phase diagram.  The energy of this pseudogap seems comparable with the superconducting gap, as seen in STM \cite{Renner:1998} , and ARPES experiments have measured its angular dependence, similar also to the superconducting gap \cite{Marshall:1996}.

The different scenarios which attempt to account for this phenomenon may be separated into two classes.  The first class of theories attribute this feature to a "precursor pairing".   Since the phase stiffness is low in these compounds, pairs may form at the pseudogap temperature ($T*$), well above $T_c$  without acquiring long range phase coherence, and then condense at $T_c$ \cite{Emery:1995}.   In some recent experiments, the observation of a large Nernst signal above $T_C$ has been attributed to the existence of vortices, seeming to plead in favor of this scenario.  However, no theory has calculated so far the Nernst signal expected in this case and it seems that gaussian fluctuations may indeed account for this effect \cite{Ussishkin:2002}. 
The second class attributes this pseudogap phase to a competing hidden order which is associated to a symmetry breaking in the normal state (at $T^*$) such as, for example, anti-ferromagnetic fluctuations,  current loops in the Cu-O plaquettes or 1D-stripes. For instance, in the  current-loop model, time-reversal symmetry and inversion symmetry are broken below $T*$  \cite{Varma:1997,Simon:2002,Kaminski:2002,Fauque:2006}.  Moshchalkov and coworkers relate the existence of the pseudogap to the formation of 1D stripes \cite{Zaanen:1989} below $T^*$  leading to translational symmetry breaking \cite{Moshchalkov:1999,Moshchalkov:2002}.  This is supported by the fitting of the zero-field resistivity -in the metallic part of the phase diagram- of LSCO thin films by a universal law $\rho (T)=\rho_0+CTexp(-\Delta/T)$, where only $\Delta$ and $\rho_0$ depend on the doping level. $\Delta$ extracted from this fit varies with doping as expected for the pseudogap and co\"{\i}ncide with NMR data.

In order to get a better insight into physics below T*, superconducting fluctuations are of key interest. If precursor pairing occurs with exactly the same type of pairs formed first at $T*$ and then condensed at $T_C$, then conventional Ginzburg-Landau fluctuations \cite{Aslamasov:1968} are not to be expected since the amplitude of the wave-function is already defined below $T*$, and only phase fluctuations are expected.  On the contrary, these phase fluctuations should lead to the resistivity variation given by the Halperin and Nelson (HN) model based on an analogy with the Kosterlitz-Thouless 2D model\cite{Halperin:1979}. 
\be
\Delta\sigma=0.37 b^{-1}\sigma_N \sinh^2[(b\tau_c/\tau)^{1/2}]
\ee
 However, all measurements of the conductivity due to fluctuations in YBCO  or BSCCO at optimal doping seem to be in favor of conventional either 2D or 3D Aslamasov-Larkin (AL) fluctuations \cite{Balestrino:1992,Cimberle:1997,Leridon:2001} \footnote{ In any case, no Maki-Thomson contribution was found nor expected due to the gap anisotropy. }. This seems to rule out precursor pairing  as being responsible for the superconducting transition at least at optimal doping.

In the underdoped regime of the (T,x) phase diagram, the problem which arises is the choice to be made for the resistivity in the normal state.  Previous analyses have been made upon the assumption that the resistivity in  the normal state remains linear as long as no charge carrier localization is present \cite{Leridon:2001, Leridon:2002, Luo:2003, Leridon:2003} , i.e. for slightly underdoped compounds.  \footnote{For the most underdoped compounds a variable range hopping (VRH) resistivity was added to this linear resistivity.} This analysis had allowed to evidence both an Aslamasov-Larkin regime in YBCO thin films and a high-temperature power law conductivity decrease \cite{Leridon:2001}.  The $T^{-\alpha}$  variation of the paraconductivity which corresponds actually to a steeper decrease than AL regime was found to be  governed by a reduced temperature $\epsilon_0=\frac{1}{\alpha} $ which increases with underdoping.   The energies associated with $\epsilon_0$ were found to be of the order of magnitude of the pseudogap energy. Caprara et al. have shown that this decrease is exactly equivalent to a high-energy cut-off \cite{Caprara:2005}.
  A total energy cutoff had also been introduced by \cite{Carballeira:2001,Mishonov:2003}.

Other groups have analyzed the data by assuming a different shape for the normal state resistivity like a downward curve fitting the high-temperature resistivity down to an arbitrary temperature.  They also came out with an AL regime for the fluctuations \cite{Vina:2002}.  It is indeed expected that the divergence of the paraconductivity at the superconducting transition should not depend on the exact variation of the resistivity in the normal state since the variation of the latter are expected to be finite in the vicinity of $T_c$ .  This is the reason why, although  different groups use different models for the normal state, they all come out with the same divergence of the paraconductivity near Tc, for small values of $\epsilon=Ln(\frac{T}{T_c})$. 

At higher temperature however, the choice made for the normal state may become more relevant.  Caprara et al.  \cite{Caprara:2005} have shown that the range of validity of the AL fluctuations depends on the choice made for the normal state resistance and that for slightly underdoped compounds, the broader range is obtained for a linear resistivity .

These studies therefore seemed to indicate that the superconducting fluctuations do behave quite conventionally close to $T_c$ in cuprates near optimal doping, except that a high-energy cut-off is present.  
As a matter of fact, in the YBa$_2$Cu$_3$O$_{7-\delta}$ and Bi$_2$Sr$_2$CaCu$_2$O$_{8+\delta}$ analyses, the normal state resistivity could only be hypothesized since $T_C$   is too high to be suppressed by the magnetic fields typically available in the laboratory.  For strongly underdoped compounds it is known that this resistivity is no more linear in T and rather governed by some localization effects, whose nature is still under debate. Therefore the observed power-law for the high-temperature variation of the paraconductivity could be questioned in relation to the choice made for the normal state resistivity.

In order to address the behavior of the fluctuations without any assumption for the normal state, we have performed high-field measurements of the resistivity of two underdoped La$_{2-x}$Sr$_x$CuO$_4$ thin films.  This material is interesting for two reasons: firstly because of its superconducting 2D-character and secondly because of its relatively low $T_C$ allowing the complete suppression of superconductivity by using pulsed fields \cite{Vanacken:2001}.  In 2D, the contribution to resistivity of the Aslamasov-Larkin fluctuations, or more precisely here of the  Lawrence-Doniach fluctuations \cite{Lawrence:1971}, is universal, depends only upon T and is  usually expressed as $\epsilon=Ln(\frac{T}{T_c})$.  
\be
 \Delta\sigma=\frac{e^2}{16\hbar d\epsilon}
 \ee
Once $T_C$ is known, the only remaining parameter is $d$, the spacing between the CuO planes, which is well known from the crystallographic characterization.  As opposed to the 3D AL paraconductivity which depends upon the zero temperature c-axis coherence length, for the 2D case no free parameters are left.  This makes the observation of a 2D AL paraconductivity highly irrefutable.  On the other hand,  measurements under high magnetic field (50 T) allow to determine precisely the normal state conductivity in order to subtract it from the measured conductivity.

  \begin{figure}
\begin{center} 
\includegraphics{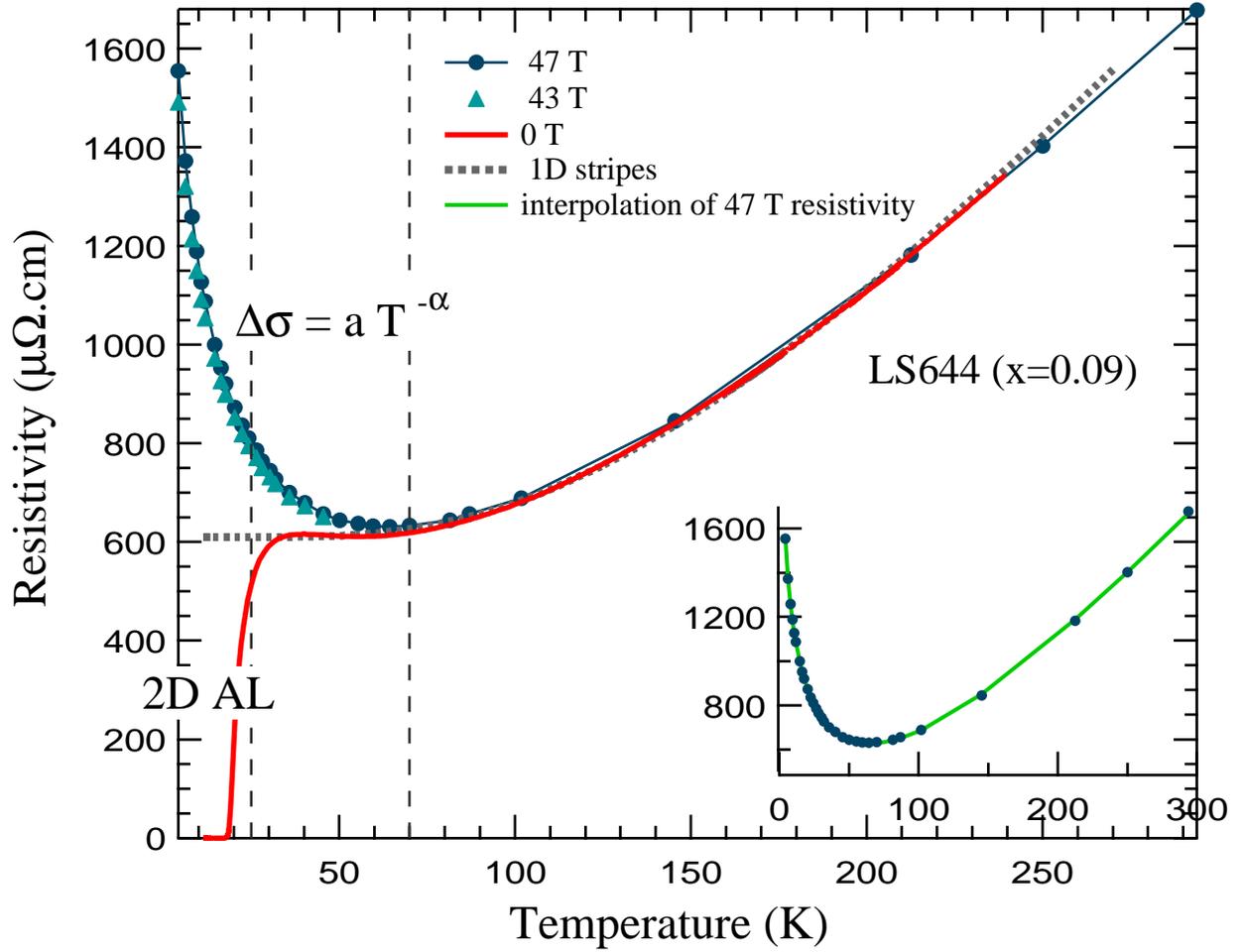}
\caption{ab-plane resistance as a function of temperature of the LSCO thin film under high magnetic field.  
The dotted line corresponds to the 1D-stripe model $\rho=\rho_0+C*T \exp{(-\Delta/T)}$, with $\rho_0 = 629 \mu \Omega .cm $, $\Delta = 256 K$ and $C = 9.38 \mu\Omega. cm. K^{-1}$.  The inset shows the 47 T data with an interpolation of the form $\rho [\mu\Omega . cm]=533.7Ln(80.3/T[K]) + 8.21T [K]$.}
\label{default}
\end{center}
\end{figure}

\section{Measurements}

The as-grown La$_{2-x}$Sr$_x$CuO$_4$ films were prepared by DC magnetron sputtering from stoichiometric targets at K.U. LeuvenÊ\cite{Wagner:2001}.Ê The transport measurements were carried out at the K.U. Leuven high field facility.Ê The reported data was obtained on thin epitaxial films of typical thickness a few hundred nm, with a patterned strip ( 1mm $\times$ 100 $ \mu$m) for four probe measurements.Ê The c-axis orientated film was mounted with $\mu_0H // c$,Ê and the current was always in the ab-plane (IÊ//Êab).Ê 
The resistivity was measured at zero magnetic field and for various intensities of the pulsed magnetic field up to  47 T.
Figure 1 shows the 0T, 43T and 47T resistivity as a function of temperature of sample LS644  whose Sr content is 0.09 and $T_C$ is 19.0 K, under different intensities of the magnetic field which was applied perpendicularly to the CuO layers. 
As can be inferred from Fig. 1, the resistivity has almost saturated between 43 T and 47 T, which allows to consider the 47 T state a reasonably good representative of the normal state. ($\frac{\rho_{47T} -\rho_{45T}}{\rho_{47T}}<2/100$). An interpolation of the 47 T resistivity is then used to obtain $\sigma_{47T}=\frac{1}{\rho_{47T}}$ and the paraconductivity is calculated as $\Delta \sigma=\sigma_{0T}-\sigma_{47T}$.  The 47T resistivity of LS644 can  actually be fitted with the function $\rho=\rho_1Ln(\frac{T_0}{T})-aT$, where $\rho_1=553.7 \mu\Omega .cm$, $T_0=80.3 K$ and $a=8.21 \mu\Omega .cm.K^{-1}$, which was used for the interpolation (see the inset in Fig.1). Such a $Ln(T)$ behavior was already observed by Ando et al.\cite{Ando:1995}.

Fig.1 also shows that above about 70 K, the 0T resistivity can be fitted with the law established for 1D resistivity in the presence of stripes: $\rho=\rho_0+C*T \exp{(-\Delta/T)}$, with $\rho_0 = 629 \mu\Omega .cm$, $\Delta = 256 K$ and $C = 9.38 \mu\Omega. cm.K^{-1}$ as was found previously \cite{Moshchalkov:2002,Weckuysen:2002}. 

$\Delta \sigma $ is then plotted as a function of $\epsilon$ on a log-log scale together with the 2D AL prediction, taking for the spacing between the CuO planes  $d=6.6\AA$ (Fig.2a and b). The agreement is quite good from $\epsilon=0.02$ to $\epsilon=0.2$, exactly the same range in $\epsilon$ where this regime was found in optimally doped $YBa_2Cu_3O_{7-\delta}$ and $Bi_2Sr_2CaCu_2O_{8+d}$  samples \cite{Cimberle:1997}. Although this range is rather narrow, it is worth noticing that, \textit{without any adjustable parameter} (except $T_c$ but this parameter is hardly adjustable \footnote{$T_C$ has been taken to be equal to 19.0 K which is between the temperature at the inflexion point (19.2 K) and the temperature of intersection between the tangent at the inflexion point and the temperature axis (18.8K), which are the two extrema allowed for $T_C$.}), \textit{both the slope and the amplitude of $\Delta \sigma $ match the AL 2D predictions within the error bars}.  These are due to the uncertainty in the sample dimensions which gives systematic error.
For comparison, the 3D AL prediction is plotted for a value of $\xi_{C0}=0.5 \AA$.  Changing the value of $\xi_{C0} $ only shifts the line vertically, so the data show no evidence of a 3D AL behavior. 
 At higher temperatures, the above-mentioned steeper decrease of the paraconductivity is observed, as can be seen in Fig. 3.
  where $\Delta \sigma$ is plotted as a function of T.  
  The power-law in T is again evidenced here  between 24K and 80 K, the normal state being straightforwardly determined.  
  The exponent $\alpha$ is found to be equal to 3.0, which gives $\epsilon_0=1/\alpha=0.33$. 
   Another sample with x=0.08 was measured at the LNCMP-Toulouse 
    high field facility under magnetic field up to 50 T.  The same high field resistivity interpolation was used to determine the normal state; the superconductive fluctuations were measured and also found consistent with 2D AL fluctuations. 
    At higher temperatures the same power law decrease of the paraconductivity in $T^{-\alpha}$ was observed with $\alpha\approx3$  (see the inset in Fig.3).

  \begin{figure}
\begin{center} 

\includegraphics{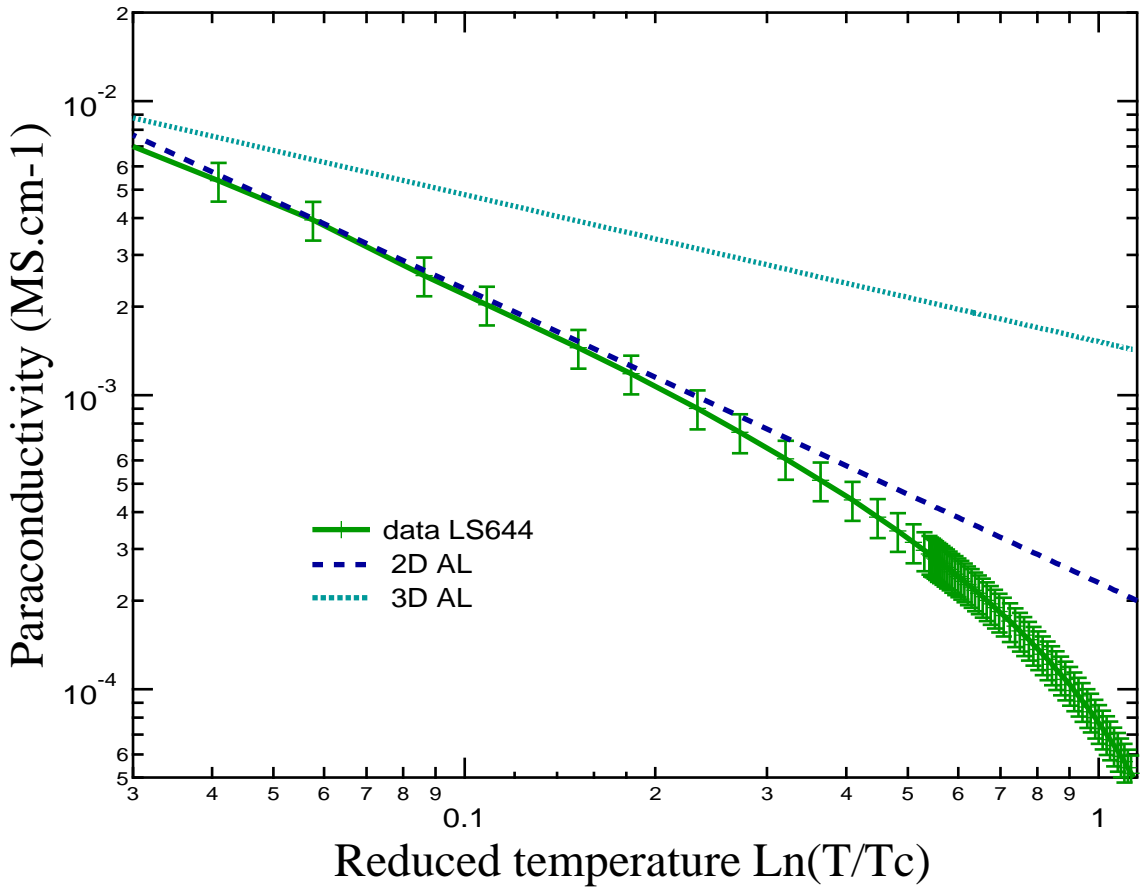}
\caption{Green markers: Paraconductivity  $ \Delta \sigma $ as a function of the reduced temperature $ \epsilon=Log(\frac{T}{T_C})$ for sample LS644.  Light blue dotted line: 3D Aslamazov-Larkin model with $\xi_{C0}=0.5\AA$. Marine blue dashed line: 2D AL model with $d = 6.6\AA$. }
\label{default}
\end{center}
\end{figure}

  \begin{figure}
\begin{center} 
\includegraphics{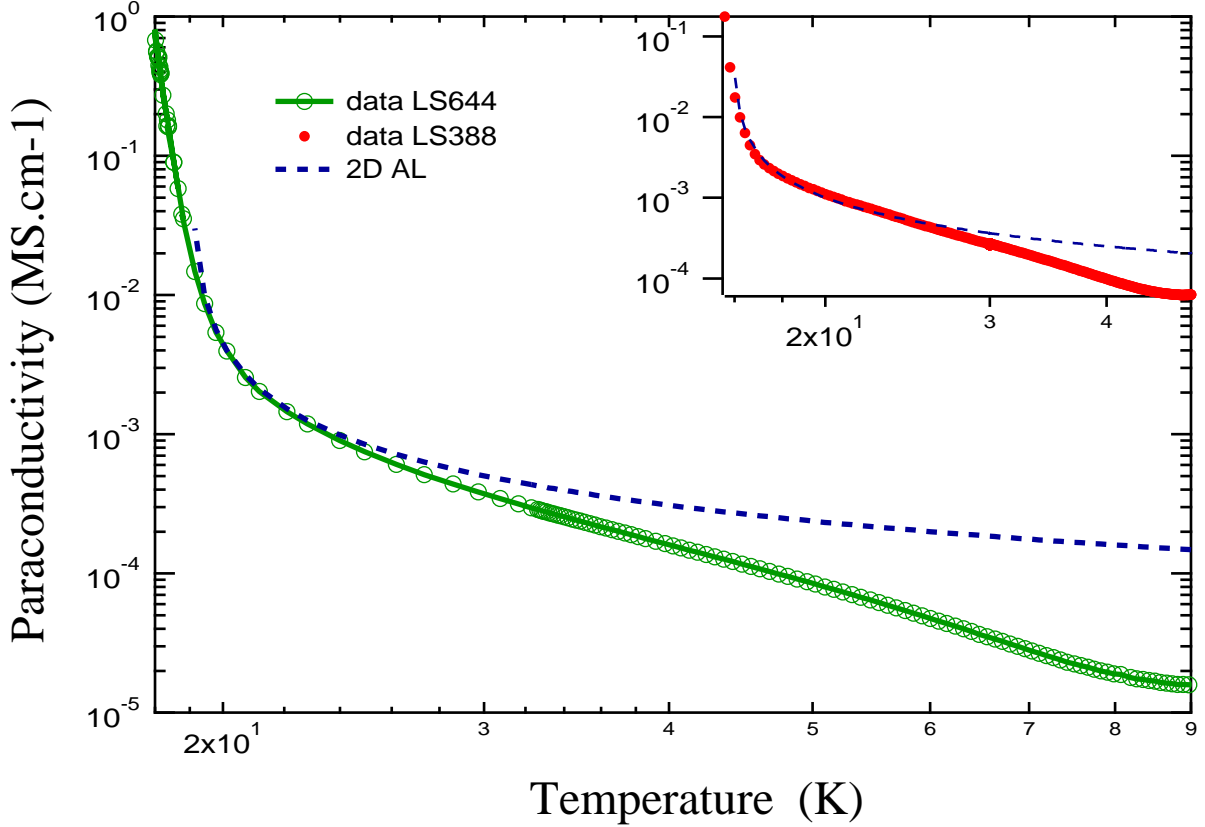}
\caption{ Paraconductivity  $ \Delta \sigma $ as a function of the temperature. Green circles: LS644; red dots LS388 (x=0.08). After the AL2D regime (blue dashed line) a linear regime is observed, which corresponds to a power-law in $T^{-\alpha}$ with the exponent $\alpha\approx3.$ }
\label{default}
\end{center}
\end{figure}

The quantitative observation of a 2D AL regime for the paraconductivity of a very underdoped LSCO compound is rather surprising.  As a matter of fact, the Fermi surface in the normal state of this material is suppressed by the opening of a pseudogap, which for that level of doping extends to almost all directions in the k-space.  The fact that, in that framework, AL predictions should remain valid is questionable, although this model does not depend explicitly on the density of states at the Fermi energy. 
Besides, a model based on phase fluctuations has been proposed for the pseudogap \cite{Emery:1995}.  In this case the fluctuation conductivity could be given by the Halperin  and Nelson (HN) expression for the conductivity in the framework of a Kosterlitz-Thouless transition \cite{Halperin:1979}, where vortices are present above $T_C$.  Here, the 50 T resistivity does not represent the "normal resistivity", because a 50T field is not sufficient to overcome a pairing energy equivalent to about 300K  .  However, one is able to reconstruct the "true normal state" resistivity from the experimental data at 0 field and the HN formula. The resistivity of the normal state would then have to be \textit{two orders of magnitude larger} than the measured resistivity (at 0T or 50T) at 200 K, with an exponential divergence at $T_C$ (since the measured resistivity is not found to diverge exponentially at $T_C$), which seems rather improbable. 

In the framework of the stripes scenario for high-Tc superconductivity, these results are a strong indication that 1D-stripes would have to encounter a 2D coupling in order to provide the superconducting ground state, whereas they would recover 1D behavior under high magnetic field.

\section{Conclusion}

Careful measurements of the resistivity of underdoped thin films of LSCO under  magnetic field up to 50 T have allowed us to extract the paraconductivity without any assumption about the normal state behavior.  The observed saturation of the resistivity with the magnetic field is an indication that the magnetoresistance of the normal state is negligible. With no adjustable parameter, the paraconductivity  quantitatively shows a  two-dimensional Aslamazov-Larkin regime near Tc and a power-law dependence at higher temperatures (typically up to 80 K).  
This behavior is in contradiction with what should be expected for preformed superconducting pairs, where a 2D Kosterlitz-Thouless behavior could be expected below $T^*$ with exponential variations in T.  Therefore these results seem to rule out preformed pairs as responsible for the superconducting transition at $T_C$ and to suggest that, quite surprisingly, the validity of the 2D AL regime of fluctuations survives the opening of a well developed pseudogap in the Fermi surface. 
Although we do not have hitherto a complete understanding of the mechanism which dampens the fluctuations at high temperature, these results may indicate a competing process between rather conventional superconducting fluctuations and the mechanism responsible for the pseudogap.  

\acknowledgments{BL gratefully acknowledges discussions with M. Grilli and S. Caprara. The work at the K.U.Leuven was supported by the FWO and the KU Leuven Research Fund GOA/2004/02 projects. B.L. acknowledges the ESF for support through the THIOX short visit grant number 1081. 

\bibliography{Mybib}

\begin{thebibliography}{30}
\expandafter\ifx\csname natexlab\endcsname\relax\def\natexlab#1{#1}\fi
\expandafter\ifx\csname bibnamefont\endcsname\relax
  \def\bibnamefont#1{#1}\fi
\expandafter\ifx\csname bibfnamefont\endcsname\relax
  \def\bibfnamefont#1{#1}\fi
\expandafter\ifx\csname citenamefont\endcsname\relax
  \def\citenamefont#1{#1}\fi
\expandafter\ifx\csname url\endcsname\relax
  \def\url#1{\texttt{#1}}\fi
\expandafter\ifx\csname urlprefix\endcsname\relax\def\urlprefix{URL }\fi
\providecommand{\bibinfo}[2]{#2}
\providecommand{\eprint}[2][]{\url{#2}}

\bibitem[{\citenamefont{Alloul et~al.}(1989)\citenamefont{Alloul, Ohno, and
  Mendels}}]{Alloul:1989}
\bibinfo{author}{\bibfnamefont{H.}~\bibnamefont{Alloul}},
  \bibinfo{author}{\bibfnamefont{T.}~\bibnamefont{Ohno}}, \bibnamefont{and}
  \bibinfo{author}{\bibfnamefont{P.}~\bibnamefont{Mendels}},
  \bibinfo{journal}{Phys. Rev. Lett.} \textbf{\bibinfo{volume}{63}},
  \bibinfo{pages}{16} (\bibinfo{year}{1989}).

\bibitem[{\citenamefont{Liang et~al.}(1996)\citenamefont{Liang, Loram, Mirza,
  Athanassopoulou, and Cooper}}]{Liang:1996}
\bibinfo{author}{\bibfnamefont{W.}~\bibnamefont{Liang}},
  \bibinfo{author}{\bibfnamefont{J.}~\bibnamefont{Loram}},
  \bibinfo{author}{\bibfnamefont{K.}~\bibnamefont{Mirza}},
  \bibinfo{author}{\bibfnamefont{N.}~\bibnamefont{Athanassopoulou}},
  \bibnamefont{and} \bibinfo{author}{\bibfnamefont{J.}~\bibnamefont{Cooper}},
  \bibinfo{journal}{Physica C} \textbf{\bibinfo{volume}{263}},
  \bibinfo{pages}{277} (\bibinfo{year}{1996}).

\bibitem[{\citenamefont{Renner et~al.}(1998)\citenamefont{Renner, Revaz,
  Genoud, Kadowaki, and Fischer}}]{Renner:1998}
\bibinfo{author}{\bibfnamefont{C.}~\bibnamefont{Renner}},
  \bibinfo{author}{\bibfnamefont{B.}~\bibnamefont{Revaz}},
  \bibinfo{author}{\bibfnamefont{J.-Y.} \bibnamefont{Genoud}},
  \bibinfo{author}{\bibfnamefont{K.}~\bibnamefont{Kadowaki}}, \bibnamefont{and}
  \bibinfo{author}{\bibfnamefont{O.}~\bibnamefont{Fischer}},
  \bibinfo{journal}{Phys. Rev. Lett.} \textbf{\bibinfo{volume}{80}},
  \bibinfo{pages}{149} (\bibinfo{year}{1998}).

\bibitem[{\citenamefont{Marshall et~al.}(1996)\citenamefont{Marshall, Dessau,
  Loeser, Park, Matsuura, Eckstein, Bozovic, Fournier, Kapitulnik, Spicer
  et~al.}}]{Marshall:1996}
\bibinfo{author}{\bibfnamefont{D.~S.} \bibnamefont{Marshall}},
  \bibinfo{author}{\bibfnamefont{D.~S.} \bibnamefont{Dessau}},
  \bibinfo{author}{\bibfnamefont{A.~G.} \bibnamefont{Loeser}},
  \bibinfo{author}{\bibfnamefont{C.-H.} \bibnamefont{Park}},
  \bibinfo{author}{\bibfnamefont{A.~Y.} \bibnamefont{Matsuura}},
  \bibinfo{author}{\bibfnamefont{J.~N.} \bibnamefont{Eckstein}},
  \bibinfo{author}{\bibfnamefont{I.}~\bibnamefont{Bozovic}},
  \bibinfo{author}{\bibfnamefont{P.}~\bibnamefont{Fournier}},
  \bibinfo{author}{\bibfnamefont{A.}~\bibnamefont{Kapitulnik}},
  \bibinfo{author}{\bibfnamefont{W.~E.} \bibnamefont{Spicer}},
  \bibnamefont{et~al.}, \bibinfo{journal}{Phys. Rev. Lett.}
  \textbf{\bibinfo{volume}{76}}, \bibinfo{pages}{4841} (\bibinfo{year}{1996}).

\bibitem[{\citenamefont{Emery and Kivelson}(1995)}]{Emery:1995}
\bibinfo{author}{\bibfnamefont{V.}~\bibnamefont{Emery}} \bibnamefont{and}
  \bibinfo{author}{\bibfnamefont{S.}~\bibnamefont{Kivelson}},
  \bibinfo{journal}{Nature} \textbf{\bibinfo{volume}{374}},
  \bibinfo{pages}{434} (\bibinfo{year}{1995}).

\bibitem[{\citenamefont{Ussishkin et~al.}(2002)\citenamefont{Ussishkin, Sondhi,
  and Huse}}]{Ussishkin:2002}
\bibinfo{author}{\bibfnamefont{I.}~\bibnamefont{Ussishkin}},
  \bibinfo{author}{\bibfnamefont{S.}~\bibnamefont{Sondhi}}, \bibnamefont{and}
  \bibinfo{author}{\bibfnamefont{D.}~\bibnamefont{Huse}},
  \bibinfo{journal}{Phys. Rev. Letters} \textbf{\bibinfo{volume}{89}},
  \bibinfo{pages}{287001} (\bibinfo{year}{2002}).

\bibitem[{\citenamefont{Simon and Varma}(2002)}]{Simon:2002}
\bibinfo{author}{\bibfnamefont{M.~E.} \bibnamefont{Simon}} \bibnamefont{and}
  \bibinfo{author}{\bibfnamefont{C.}~\bibnamefont{Varma}},
  \bibinfo{journal}{Phys. Rev. Letters} \textbf{\bibinfo{volume}{89}},
  \bibinfo{pages}{247003} (\bibinfo{year}{2002}).

\bibitem[{\citenamefont{Kaminski et~al.}(2002)\citenamefont{Kaminski,
  Rosenkranz, Fretwell, Campuzano, Li, Raffy, Cullen, You, Olsonk, Varma
  et~al.}}]{Kaminski:2002}
\bibinfo{author}{\bibfnamefont{A.}~\bibnamefont{Kaminski}},
  \bibinfo{author}{\bibfnamefont{S.}~\bibnamefont{Rosenkranz}},
  \bibinfo{author}{\bibfnamefont{H.~M.} \bibnamefont{Fretwell}},
  \bibinfo{author}{\bibfnamefont{J.~C.} \bibnamefont{Campuzano}},
  \bibinfo{author}{\bibfnamefont{Z.}~\bibnamefont{Li}},
  \bibinfo{author}{\bibfnamefont{H.}~\bibnamefont{Raffy}},
  \bibinfo{author}{\bibfnamefont{W.~G.} \bibnamefont{Cullen}},
  \bibinfo{author}{\bibfnamefont{H.}~\bibnamefont{You}},
  \bibinfo{author}{\bibfnamefont{C.~G.} \bibnamefont{Olsonk}},
  \bibinfo{author}{\bibfnamefont{C.~M.} \bibnamefont{Varma}},
  \bibnamefont{et~al.}, \bibinfo{journal}{Nature} \textbf{\bibinfo{volume}{\bf
  416}}, \bibinfo{pages}{610} (\bibinfo{year}{2002}).

\bibitem[{\citenamefont{Fauque et~al.}(2006)\citenamefont{Fauque, Sidis,
  Hinkov, Pailhes, Lin, Chaud, and Bourges}}]{Fauque:2006}
\bibinfo{author}{\bibfnamefont{B.}~\bibnamefont{Fauque}},
  \bibinfo{author}{\bibfnamefont{Y.}~\bibnamefont{Sidis}},
  \bibinfo{author}{\bibfnamefont{V.}~\bibnamefont{Hinkov}},
  \bibinfo{author}{\bibfnamefont{S.}~\bibnamefont{Pailhes}},
  \bibinfo{author}{\bibfnamefont{C.~T.} \bibnamefont{Lin}},
  \bibinfo{author}{\bibfnamefont{X.}~\bibnamefont{Chaud}}, \bibnamefont{and}
  \bibinfo{author}{\bibfnamefont{P.}~\bibnamefont{Bourges}},
  \bibinfo{journal}{Phys. Rev. Letters} \textbf{\bibinfo{volume}{\bf96}},
  \bibinfo{pages}{197001} (\bibinfo{year}{2006}).

\bibitem[{\citenamefont{C.M.Varma}(1997)}]{Varma:1997}
\bibinfo{author}{\bibnamefont{C.M.Varma}}, \bibinfo{journal}{Phys. Rev. B}
  \textbf{\bibinfo{volume}{55}}, \bibinfo{pages}{14554} (\bibinfo{year}{1997}).

\bibitem[{\citenamefont{Zaanen and Gunnarsson}(1989)}]{Zaanen:1989}
\bibinfo{author}{\bibfnamefont{J.}~\bibnamefont{Zaanen}} \bibnamefont{and}
  \bibinfo{author}{\bibfnamefont{O.}~\bibnamefont{Gunnarsson}},
  \bibinfo{journal}{Phys. Rev. B} \textbf{\bibinfo{volume}{40}},
  \bibinfo{pages}{7391} (\bibinfo{year}{1989}).

\bibitem[{\citenamefont{Moshchalkov et~al.}(1999)\citenamefont{Moshchalkov,
  Vanacken, and Trappeniers}}]{Moshchalkov:1999}
\bibinfo{author}{\bibfnamefont{V.~V.} \bibnamefont{Moshchalkov}},
  \bibinfo{author}{\bibfnamefont{J.}~\bibnamefont{Vanacken}}, \bibnamefont{and}
  \bibinfo{author}{\bibfnamefont{L.}~\bibnamefont{Trappeniers}},
  \bibinfo{journal}{Europhys. Lett.} \textbf{\bibinfo{volume}{46}},
  \bibinfo{pages}{75} (\bibinfo{year}{1999}).

\bibitem[{\citenamefont{Moshchalkov et~al.}(2002)\citenamefont{Moshchalkov,
  Vanacken, and Trappeniers}}]{Moshchalkov:2002}
\bibinfo{author}{\bibfnamefont{V.}~\bibnamefont{Moshchalkov}},
  \bibinfo{author}{\bibfnamefont{J.}~\bibnamefont{Vanacken}}, \bibnamefont{and}
  \bibinfo{author}{\bibfnamefont{L.}~\bibnamefont{Trappeniers}},
  \bibinfo{journal}{Phys. Rev. B} \textbf{\bibinfo{volume}{64}},
  \bibinfo{pages}{214504} (\bibinfo{year}{2002}).

\bibitem[{\citenamefont{Aslamasov and Larkin}(1968)}]{Aslamasov:1968}
\bibinfo{author}{\bibfnamefont{L.}~\bibnamefont{Aslamasov}} \bibnamefont{and}
  \bibinfo{author}{\bibfnamefont{A.}~\bibnamefont{Larkin}},
  \bibinfo{journal}{Phys. Letters A} \textbf{\bibinfo{volume}{26}},
  \bibinfo{pages}{238} (\bibinfo{year}{1968}).

\bibitem[{\citenamefont{Halperin and Nelson}(1979)}]{Halperin:1979}
\bibinfo{author}{\bibfnamefont{B.}~\bibnamefont{Halperin}} \bibnamefont{and}
  \bibinfo{author}{\bibfnamefont{D.}~\bibnamefont{Nelson}},
  \bibinfo{journal}{J. of Low Temp. Phys.} \textbf{\bibinfo{volume}{\bf36}},
  \bibinfo{pages}{599} (\bibinfo{year}{1979}).

\bibitem[{\citenamefont{Balestrino et~al.}(1992)\citenamefont{Balestrino,
  Marinelli, Milani, Reggiano, Vaglio, and Varlamov}}]{Balestrino:1992}
\bibinfo{author}{\bibfnamefont{G.}~\bibnamefont{Balestrino}},
  \bibinfo{author}{\bibfnamefont{M.}~\bibnamefont{Marinelli}},
  \bibinfo{author}{\bibfnamefont{E.}~\bibnamefont{Milani}},
  \bibinfo{author}{\bibfnamefont{L.}~\bibnamefont{Reggiano}},
  \bibinfo{author}{\bibfnamefont{R.}~\bibnamefont{Vaglio}}, \bibnamefont{and}
  \bibinfo{author}{\bibfnamefont{A.~A.} \bibnamefont{Varlamov}},
  \bibinfo{journal}{Phys. Rev. B} \textbf{\bibinfo{volume}{46}},
  \bibinfo{pages}{14919} (\bibinfo{year}{1992}).

\bibitem[{\citenamefont{Cimberle et~al.}(1997)\citenamefont{Cimberle,
  Ferdeghini, Giannini, Marre, Putti, Siri, Federici, and
  Varlamov}}]{Cimberle:1997}
\bibinfo{author}{\bibfnamefont{M.~R.} \bibnamefont{Cimberle}},
  \bibinfo{author}{\bibfnamefont{C.}~\bibnamefont{Ferdeghini}},
  \bibinfo{author}{\bibfnamefont{E.}~\bibnamefont{Giannini}},
  \bibinfo{author}{\bibfnamefont{D.}~\bibnamefont{Marre}},
  \bibinfo{author}{\bibfnamefont{M.}~\bibnamefont{Putti}},
  \bibinfo{author}{\bibfnamefont{A.}~\bibnamefont{Siri}},
  \bibinfo{author}{\bibfnamefont{F.}~\bibnamefont{Federici}}, \bibnamefont{and}
  \bibinfo{author}{\bibfnamefont{A.}~\bibnamefont{Varlamov}},
  \bibinfo{journal}{Phys. Rev. B} \textbf{\bibinfo{volume}{55}},
  \bibinfo{pages}{R14745} (\bibinfo{year}{1997}).

\bibitem[{\citenamefont{Leridon et~al.}(2001)\citenamefont{Leridon, D\'efossez,
  Dumont, Lesueur, and Contour}}]{Leridon:2001}
\bibinfo{author}{\bibfnamefont{B.}~\bibnamefont{Leridon}},
  \bibinfo{author}{\bibfnamefont{A.}~\bibnamefont{D\'efossez}},
  \bibinfo{author}{\bibfnamefont{J.}~\bibnamefont{Dumont}},
  \bibinfo{author}{\bibfnamefont{J.}~\bibnamefont{Lesueur}}, \bibnamefont{and}
  \bibinfo{author}{\bibfnamefont{J.~P.} \bibnamefont{Contour}},
  \bibinfo{journal}{Phys. Rev. Lett.} \textbf{\bibinfo{volume}{87}},
  \bibinfo{pages}{197007} (\bibinfo{year}{2001}).

\bibitem[{\citenamefont{Leridon et~al.}(2002)\citenamefont{Leridon, Moragu\`es,
  Lesueur, D\'efossez, Dumont, and Contour}}]{Leridon:2002}
\bibinfo{author}{\bibfnamefont{B.}~\bibnamefont{Leridon}},
  \bibinfo{author}{\bibfnamefont{M.}~\bibnamefont{Moragu\`es}},
  \bibinfo{author}{\bibfnamefont{J.}~\bibnamefont{Lesueur}},
  \bibinfo{author}{\bibfnamefont{A.}~\bibnamefont{D\'efossez}},
  \bibinfo{author}{\bibfnamefont{J.}~\bibnamefont{Dumont}}, \bibnamefont{and}
  \bibinfo{author}{\bibfnamefont{J.~P.} \bibnamefont{Contour}},
  \bibinfo{journal}{Journal of Superconductivity: Incorporating Novel
  Magnetism} \textbf{\bibinfo{volume}{15}}, \bibinfo{pages}{409}
  (\bibinfo{year}{2002}).

\bibitem[{\citenamefont{Luo et~al.}(2003)\citenamefont{Luo, andJ. Y.~Lin, H.Wu,
  Uen, and Gou}}]{Luo:2003}
\bibinfo{author}{\bibfnamefont{C.}~\bibnamefont{Luo}},
  \bibinfo{author}{\bibfnamefont{J.~J.} \bibnamefont{andJ. Y.~Lin}},
  \bibinfo{author}{\bibfnamefont{K.}~\bibnamefont{H.Wu}},
  \bibinfo{author}{\bibfnamefont{T.}~\bibnamefont{Uen}}, \bibnamefont{and}
  \bibinfo{author}{\bibfnamefont{Y.~S.} \bibnamefont{Gou}},
  \bibinfo{journal}{Phys. Rev. Lett.} \textbf{\bibinfo{volume}{90}},
  \bibinfo{pages}{179703} (\bibinfo{year}{2003}).

\bibitem[{\citenamefont{Leridon et~al.}(2003)\citenamefont{Leridon, D\'efossez,
  Dumont, Lesueur, and Contour}}]{Leridon:2003}
\bibinfo{author}{\bibfnamefont{B.}~\bibnamefont{Leridon}},
  \bibinfo{author}{\bibfnamefont{A.}~\bibnamefont{D\'efossez}},
  \bibinfo{author}{\bibfnamefont{J.}~\bibnamefont{Dumont}},
  \bibinfo{author}{\bibfnamefont{J.}~\bibnamefont{Lesueur}}, \bibnamefont{and}
  \bibinfo{author}{\bibfnamefont{J.~P.} \bibnamefont{Contour}},
  \bibinfo{journal}{Phys. Rev. Lett.} \textbf{\bibinfo{volume}{90}},
  \bibinfo{pages}{179704} (\bibinfo{year}{2003}).

\bibitem[{\citenamefont{Caprara et~al.}(2005)\citenamefont{Caprara, Grilli,
  Leridon, and Lesueur}}]{Caprara:2005}
\bibinfo{author}{\bibfnamefont{S.}~\bibnamefont{Caprara}},
  \bibinfo{author}{\bibfnamefont{M.}~\bibnamefont{Grilli}},
  \bibinfo{author}{\bibfnamefont{B.}~\bibnamefont{Leridon}}, \bibnamefont{and}
  \bibinfo{author}{\bibfnamefont{J.}~\bibnamefont{Lesueur}},
  \bibinfo{journal}{Phys. Rev. B} \textbf{\bibinfo{volume}{72}},
  \bibinfo{pages}{104509} (\bibinfo{year}{2005}).

\bibitem[{\citenamefont{Carballeira et~al.}(2001)\citenamefont{Carballeira,
  Curras, {n}a, Veira, Ramallo, and Vidal}}]{Carballeira:2001}
\bibinfo{author}{\bibfnamefont{C.}~\bibnamefont{Carballeira}},
  \bibinfo{author}{\bibfnamefont{S.~R.} \bibnamefont{Curras}},
  \bibinfo{author}{\bibfnamefont{J.~V.} \bibnamefont{{n}a}},
  \bibinfo{author}{\bibfnamefont{J.~A.} \bibnamefont{Veira}},
  \bibinfo{author}{\bibfnamefont{M.~V.} \bibnamefont{Ramallo}},
  \bibnamefont{and} \bibinfo{author}{\bibfnamefont{F.}~\bibnamefont{Vidal}},
  \bibinfo{journal}{Phys. Rev. B} \textbf{\bibinfo{volume}{63}},
  \bibinfo{pages}{144515} (\bibinfo{year}{2001}).

\bibitem[{\citenamefont{Mishonov et~al.}(2003)\citenamefont{Mishonov, Pachov,
  Genchev, Atanasova, and Damianov}}]{Mishonov:2003}
\bibinfo{author}{\bibfnamefont{T.~M.} \bibnamefont{Mishonov}},
  \bibinfo{author}{\bibfnamefont{G.~V.} \bibnamefont{Pachov}},
  \bibinfo{author}{\bibfnamefont{I.~N.} \bibnamefont{Genchev}},
  \bibinfo{author}{\bibfnamefont{L.~A.} \bibnamefont{Atanasova}},
  \bibnamefont{and} \bibinfo{author}{\bibfnamefont{D.~C.}
  \bibnamefont{Damianov}}, \bibinfo{journal}{Phys. Rev. B}
  \textbf{\bibinfo{volume}{68}}, \bibinfo{pages}{054525}
  (\bibinfo{year}{2003}).

\bibitem[{\citenamefont{Vina et~al.}(2002)\citenamefont{Vina, Campa,
  Carballeira, Curra«s, Maignan, Ramallo, Rasines, Veira, Wagner, and
  Vidal}}]{Vina:2002}
\bibinfo{author}{\bibfnamefont{J.}~\bibnamefont{Vina}},
  \bibinfo{author}{\bibfnamefont{J.~A.} \bibnamefont{Campa}},
  \bibinfo{author}{\bibfnamefont{C.}~\bibnamefont{Carballeira}},
  \bibinfo{author}{\bibfnamefont{S.~R.} \bibnamefont{Curra«s}},
  \bibinfo{author}{\bibfnamefont{A.}~\bibnamefont{Maignan}},
  \bibinfo{author}{\bibfnamefont{M.~V.} \bibnamefont{Ramallo}},
  \bibinfo{author}{\bibfnamefont{I.}~\bibnamefont{Rasines}},
  \bibinfo{author}{\bibfnamefont{J.~A.} \bibnamefont{Veira}},
  \bibinfo{author}{\bibfnamefont{P.}~\bibnamefont{Wagner}}, \bibnamefont{and}
  \bibinfo{author}{\bibfnamefont{F.}~\bibnamefont{Vidal}},
  \bibinfo{journal}{Phys. Rev. B} \textbf{\bibinfo{volume}{65}},
  \bibinfo{pages}{212509} (\bibinfo{year}{2002}).

\bibitem[{\citenamefont{J.Vanacken}(2001)}]{Vanacken:2001}
\bibinfo{author}{\bibnamefont{J.Vanacken}}, \bibinfo{journal}{Physica B}
  \textbf{\bibinfo{volume}{294}}, \bibinfo{pages}{591} (\bibinfo{year}{2001}).

\bibitem[{\citenamefont{Lawrence and Doniach}(1971)}]{Lawrence:1971}
\bibinfo{author}{\bibfnamefont{W.}~\bibnamefont{Lawrence}} \bibnamefont{and}
  \bibinfo{author}{\bibfnamefont{S.}~\bibnamefont{Doniach}},
  \bibinfo{journal}{Proc. 12th Int. Conf. on Low Temp. Phys.} p.
  \bibinfo{pages}{361} (\bibinfo{year}{1971}).

\bibitem[{\citenamefont{P.Wagner et~al.}(2001)\citenamefont{P.Wagner, Ruan,
  Gordon, Vanacken, Moshchalkov, and Bruynseraede}}]{Wagner:2001}
\bibinfo{author}{\bibnamefont{P.Wagner}},
  \bibinfo{author}{\bibfnamefont{K.}~\bibnamefont{Ruan}},
  \bibinfo{author}{\bibfnamefont{I.}~\bibnamefont{Gordon}},
  \bibinfo{author}{\bibfnamefont{J.}~\bibnamefont{Vanacken}},
  \bibinfo{author}{\bibfnamefont{V.~V.} \bibnamefont{Moshchalkov}},
  \bibnamefont{and}
  \bibinfo{author}{\bibfnamefont{Y.}~\bibnamefont{Bruynseraede}},
  \bibinfo{journal}{Physica C} \textbf{\bibinfo{volume}{356}},
  \bibinfo{pages}{107} (\bibinfo{year}{2001}).

\bibitem[{\citenamefont{Ando et~al.}(1995)\citenamefont{Ando, Boebinger,
  Passner, Kimura, and Kishio}}]{Ando:1995}
\bibinfo{author}{\bibfnamefont{Y.}~\bibnamefont{Ando}},
  \bibinfo{author}{\bibfnamefont{G.}~\bibnamefont{Boebinger}},
  \bibinfo{author}{\bibfnamefont{A.}~\bibnamefont{Passner}},
  \bibinfo{author}{\bibfnamefont{T.}~\bibnamefont{Kimura}}, \bibnamefont{and}
  \bibinfo{author}{\bibfnamefont{K.}~\bibnamefont{Kishio}},
  \bibinfo{journal}{Phys. Rev. Letters} \textbf{\bibinfo{volume}{75}},
  \bibinfo{pages}{25} (\bibinfo{year}{1995}).

\bibitem[{\citenamefont{Weckhuysen et~al.}(2002)\citenamefont{Weckhuysen,
  Vanacken, Wagner, and Moshchalkov}}]{Weckuysen:2002}
\bibinfo{author}{\bibfnamefont{L.}~\bibnamefont{Weckhuysen}},
  \bibinfo{author}{\bibfnamefont{J.}~\bibnamefont{Vanacken}},
  \bibinfo{author}{\bibfnamefont{P.}~\bibnamefont{Wagner}}, \bibnamefont{and}
  \bibinfo{author}{\bibfnamefont{V.}~\bibnamefont{Moshchalkov}},
  \bibinfo{journal}{Euro. Phys. J. B} \textbf{\bibinfo{volume}{29}},
  \bibinfo{pages}{369} (\bibinfo{year}{2002}).

\end{thebibliography}

\end{document}